\documentclass[runningheads]{llncs}
\usepackage{graphicx}
\usepackage{amsmath}
\usepackage{amssymb}
\usepackage{romanbar}
\usepackage{draftwatermark}
\SetWatermarkText{preprint}
\SetWatermarkScale{5}
\newcounter{deficit_counter}
\newcommand{\deficitbox}[1]{
    \addtocounter{deficit_counter}{1}
    \setlength\fboxsep{0.2em}
    \begin{center}
        \fbox{\parbox{0.96\linewidth}{
            \textbf{\textit{Challenge \Roman{deficit_counter}}}: #1
        }}
        \vspace{0.5em}
    \end{center}
}

\begin{document}
\title{A Roadmap for Automating the Selection of Quantum Computers for Quantum Algorithms}
\titlerunning{Selection of Quantum Computers for Quantum Algorithms}
%
\author{Marie Salm\orcidID{0000-0002-2180-250X} \and
Johanna Barzen\orcidID{0000-0001-8397-7973} \and \\
Uwe Breitenb\"ucher\orcidID{0000-0002-8816-5541} \and
Frank Leymann\orcidID{0000-0002-9123-259X} \and
\\Benjamin Weder\orcidID{0000-0002-6761-6243} \and
Karoline Wild\orcidID{0000-0001-7803-6386}}
\authorrunning{M. Salm et al.}
%
\institute{Institute of Architecture of Application Systems, University of Stuttgart, \\Universitätsstraße 38, Stuttgart, Germany
\email{\{salm,barzen,breitenbuecher,leymann,weder,wild\}@iaas.uni-stuttgart.de}}
\maketitle              
\begin{abstract}
Quantum computing can enable a variety of breakthroughs in research and industry in the future.
Although some quantum algorithms already exist that show a theoretical speedup compared to the best known classical algorithms, the implementation and execution of these algorithms come with several challenges.
The input data determines, e.g., the required number of qubits and gates of a quantum algorithm.
An algorithm implementation also depends on the used Software Development Kit which restricts the set of usable quantum computers.
Because of the limited capabilities of current quantum computers, choosing an appropriate one to execute a certain implementation for a given input is a difficult challenge that requires immense mathematical knowledge about the implemented quantum algorithm as well as technical knowledge about the used Software Development Kits.
Thus, we present a roadmap for the automated analysis and selection of implementations of a certain quantum algorithm and appropriate quantum computers that can execute the selected implementation with the given input data.

\keywords{Quantum Computing \and Quantum Algorithms \and Hardware Selection \and Implementation Selection \and Decision Support.}
\end{abstract}
\section{Introduction}
Quantum computing is a promising field that may in future enable breakthroughs in various areas such as computer science, physics, and chemistry~\cite{NAP25196}.
The unique characteristics of quantum mechanics, such as \textit{superposition} and \textit{entanglement}, are the reasons why quantum computing is more powerful than classical computing for certain problems~\cite{Arute2019,Preskill2018quantumcomputingin,rieffelgentle}.
In fact, some quantum algorithms already exist that show a theoretical speedup over their best known classical counterparts.
For example, the \textit{Shor} algorithm provides an exponential speedup in factorizing numbers~\cite{Shor_1997}.
With a large enough quantum computer, this algorithm could break cryptosystems such as the commonly used RSA~\cite{Preskill2018quantumcomputingin}.

However, there are several challenges regarding the execution of quantum algorithms, which are considered in the following.
There exists a multitude of different implementations for quantum algorithms that are only applicable to certain input data, e.g., in terms of the number of qubits required for its encoding, which we refer to as \textit{input size}.
These implementations differ from each other in various aspects, e.g., the required number of qubits and operations~\cite{haener2017factoring}.
Thereby, both numbers often depend on the input data.
This means that the input data influences whether or not a particular quantum algorithm implementation is executable on a certain quantum computer:
If the number of required qubits or operations is higher than the number of provided qubits or the decoherence time of the quantum computer, the implementation with the given input cannot be executed on this machine.
Also error rates, fidelity, and qubit connectivity of a quantum computer play an important role in the decision.

In addition, there is currently no accepted common quantum programming language~\cite{leymann2020quantum}.
As a result, most quantum computer vendors have their proprietary \textit{Software Development Kit (SDK)} for developing and executing implementations on their quantum computers~\cite{LaRose2019}.
However, this tightly couples the implementation of a quantum algorithm to a certain kind of quantum computer.
As a result, choosing an implementation for a quantum algorithm to be executed for given input data as well as selecting an appropriate quantum computer is a multi-dimensional challenge that requires immense mathematical knowledge about the implemented algorithm as well as technical knowledge about the used SDKs. 
Hence, \textit{(i) the selection of a suitable implementation of a quantum algorithm for a specific input} and \textit{(ii) the selection of a quantum computer with, e.g., enough qubits and decoherence time} is currently one of the main problems when quantum computing is used in practice.

In this paper, we present a roadmap for a \textit{NISQ Analyzer} that analyzes and selects \textit{(i) an appropriate implementation} and \textit{(ii) suitable quantum hardware depending on the input data for a chosen quantum algorithm}.
The approach is based on defining selection criteria for each implementation described as first-order logic rules.
Thereby, we take into account the number of required qubits of the implementation and the number of provided qubits of eligible quantum computers, while vendor-specific SDKs are also heeded.
In addition, the number of operations of an implementation is determined and the corresponding decoherence time of the different quantum computers are considered.
To determine the number of qubits and operations of an implementation, hardware-specific transpilers provided by the vendors are used.
The NISQ Analyzer is designed as a plug-in based system, such that additional criteria, e.g., error rates, fidelity, or qubit connectivity, can be added in the future. 
\section{Background, Challenges and Problem Statement}
In this section, we introduce the fundamentals and current challenges when using quantum computers during the \textit{Noisy Intermediate-Scale Quantum} \textit{(NISQ)} era~\cite{Preskill2018quantumcomputingin}. 
Afterward, quantum algorithms and the current state of their implementations are presented.
Finally, we formulate the problem statement and the resulting research question of this paper.

\subsection{Quantum Computers and NISQ}
Instead of calculating with classical bits, quantum computers are calculating based on so-called \textit{qubits}~\cite{nielsenchuang}.
As classical bits can only be in one of the two states 0 or 1, qubits can be in both states at the same time~\cite{nielsenchuang,rieffelgentle}. 
The state of a qubit is represented as a unit vector in a two-dimensional complex vector space, and operators that are applied to these vectors are \textit{unitary} matrices~\cite{nielsenchuang}.
Qubits interact with their environment, and thus, their states are only stable for a certain time, called \textit{decoherence time}~\cite{error,nielsenchuang,rieffelgentle}.
The required operations have to be applied in this time frame to obtain proper results from computations.
Furthermore, different quantum computing models exist, e.g., \textit{one-way}~\cite{measurementbased}, \textit{adiabatic}~\cite{aharonov2008adiabatic}, and \textit{gate-based}~\cite{nielsenchuang}.
In this paper, we only consider the gate-base quantum computing model, as many of the existing quantum computers, e.g., from IBM\footnote{https://quantum-computing.ibm.com} and Rigetti\footnote{https://www.rigetti.com}, are based on this model~\cite{LaRose2019}.
In this model, unitary operations are represented as \textit{gates}, combined with qubits and measurements they form a \textit{quantum circuit}~\cite{nielsenchuang}.
Such quantum circuits are gate-based representations of \textit{quantum algorithms}.
The number of gate collections to be executed sequentially is defined as the \textit{depth} of a quantum circuit.
Within such a collection, called \textit{layer}, gates are performed in parallel.
The number of qubits is defined as the \textit{width} of the circuit.
Both properties determine the required number of qubits and the stable execution time of a suitable quantum computer.

Each quantum computer has a set of physically implemented gates~\cite{rieffelgentle}.
However, the sets of implemented gates differ from quantum computer to quantum computer.
Thus, to create quantum circuits for specific problems, gates that are not implemented on the specific quantum computer must be realized by a combination of available gates~\cite{leymann2020quantum}.
This is done by the hardware-specific transpiler of the vendor.
Therefore, the transpiler maps the gates and qubits of the circuit to the gate sets and qubits of the regarded quantum computers.
The resulting transpiled circuit may have a different depth and width than the general circuit.
Especially the depth can differ greatly between different quantum computers.

Today's quantum computers only have a small number of qubits and short decoherence times~\cite{Tannu2018}.
Further, high error rates limit the number of operations that can be executed on these quantum computers before the propagated error make the computation too erroneous.
However, it is assumed that quantum computers will have up to a few hundred qubits and can perform thousands of operations reliably soon~\cite{Preskill2018quantumcomputingin}.
But the limitation is that these qubits will still be error-prone because for the correction of such errors many more qubits are needed~\cite{NAP25196,Preskill2018quantumcomputingin}.
Thus, these quantum computers are also called NISQ machines.

\deficitbox{\textit{There are a variety of quantum computers that are different regarding their number of qubits, their decoherence time, and their set of physically implemented gates. Therefore, there is serious heterogeneity of available quantum computers, and not every implementation can be executed on every quantum computer.}}

\subsection{Quantum Algorithms and Implementations}\label{algoimpls}
Many quantum algorithms show a theoretical speedup over their best known classical counterparts. 
The number of required qubits and operations for the execution of quantum algorithms often depends on the input data.
E.g., the Shor algorithm requires $2n$ qubits for the integer $N$ with a binary size of $n$ to be factorized~\cite{haener2017factoring}.
For some implementations, additional qubits are required for executing the algorithm.
E.g., \textit{QPE}~\cite{OBrien2019}, which can be applied to compute the eigenvalues of a unitary matrix, needs in many of the existing implementations additional qubits which define the precision of the result.
There are also implementations that can only process a limited input size, e.g., an implementation of Shor that can only factorize up to 15, but may require fewer operations than general, unlimited implementations.\footnote{https://quantum-circuit.com}
Thus, selecting an appropriate quantum computer to execute a certain quantum algorithm not only depends on the mathematics of the algorithm itself but also on the physical requirements of its implementations.

In addition, current implementations of quantum algorithms are tightly coupled to the SDKs they are developed with.
Companies like IBM and Rigetti offer their proprietary SDKs for their quantum computers, called \textit{Qiskit\footnote{https://qiskit.org}} and \textit{Forest\footnote{http://docs.rigetti.com/en/stable/}}, respectively.
There are also other SDKs that support the quantum computers of multiple vendors, e.g., \textit{ProjectQ}~\cite{projectq18} or \textit{XACC}~\cite{XACC}.
Nonetheless, most of the SDKs only support quantum computers of a single vendor as backends~\cite{LaRose2019}.
Furthermore, implementations are not interchangeable between different SDKs because of their different programming languages and syntax.
As a result, the majority of the developed implementations are only executable on a certain set of quantum computers provided by a specific vendor.

\deficitbox{\textit{An implementation of a quantum algorithm implies physical requirements on a quantum computer. In addition, an implementation usually depends on the used SDK.}}

\subsection{Problem Statement}
In this section, we summarize the challenges presented before and formulate the research question regarding the selection of quantum computers capable of executing a quantum algorithm for certain input data.
For this purpose, the user has to consider different aspects regarding available quantum algorithm implementations and quantum computers.
First, he or she has to manually find a suitable implementation for the required quantum algorithm, which can process the desired input.
With the chosen quantum algorithm implementation the user has to select a suitable quantum computer, that can be used to execute the implementation.
Thereby, the heterogeneity of the quantum hardware, with their different qubit counts, decoherence times, and available gate sets, has to be taken into account~(\textit{\textbf{Challenge~\Romannum{1}}}).
Additionally, the mathematical and technical requirements on the quantum computer and the utilized SDK of the implementation have to be considered for the quantum computer selection~(\textit{\textbf{Challenge~\Romannum{2}}}).
Thus, the selection of implementations and suitable quantum computers requires an immense manual effort and sufficient knowledge on the user side.
Hence, the resulting research question can be formulated as follows: \textit{"How can the selection of the quantum algorithm implementation and the suitable quantum hardware be automated based on the input data of the chosen quantum algorithm?"}

\section{Approach}

\begin{figure}[t!]
	\centering
	\includegraphics[width=1\linewidth]{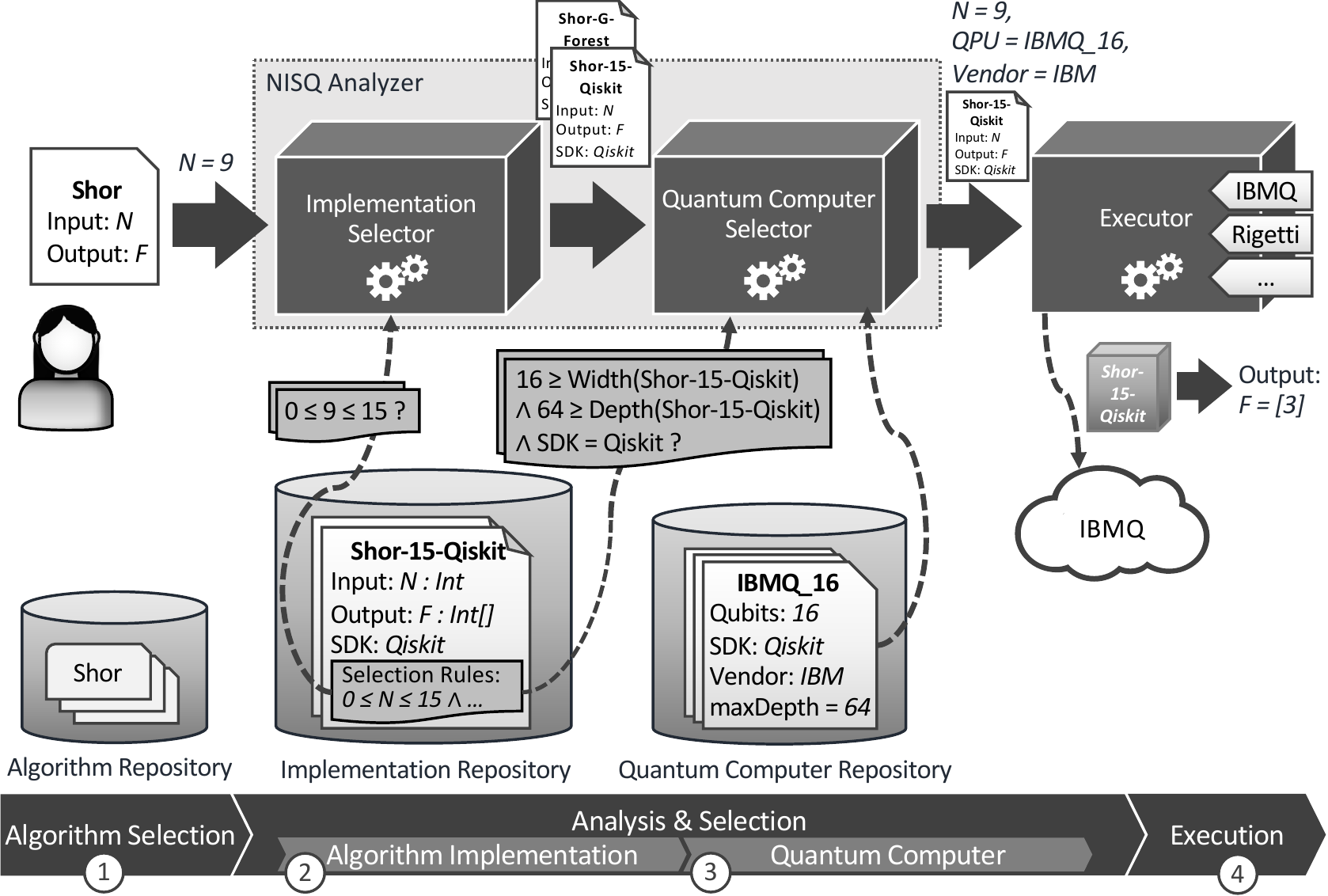}
    \caption{Approach for automated implementation and quantum hardware selection.}
    \label{fig:approach}
\end{figure}

In this section, we introduce our approach for a NISQ Analyzer, which enables an automated analysis and selection of algorithm implementations and quantum hardware depending on the chosen quantum  algorithm and specific input data.
Fig.~\ref{fig:approach} depicts an overview of the approach.

In the \textit{(1) Algorithm Selection} phase, the user has to select one of the provided quantum algorithms for solving a particular problem, e.g., the Shor algorithm as shown in Fig.~\ref{fig:approach}.
Thus, a repository with a set of different quantum algorithms is provided, as proposed by~\cite{leymann2019platform}.
After the selection of the quantum algorithm, the required input parameters of the algorithm and the corresponding implementations have to be provided, e.g., for Shor a natural number $N$.

Next, in the \textit{Analysis \& Selection} phase, first applicable implementations for the given input and then appropriate quantum computers are identified.
In the \textit{(2) Algorithm Implementation Analysis \& Selection} phase the repository with available implementations is browsed to identify implementations that can process the given input.
For this, selection rules, described by first-order logic, are attached to the implementations.
A rule describes the restrictions for the input data of the respective implementation.
As seen in Fig.~\ref{fig:approach}, the selection rule for the implementation \textit{Shor-15-Qiskit} is defined as follows: 
\begin{equation}
\begin{aligned}
y \in \mathcal{I}, \forall n \in \mathbb{N},\exists l_0, l_1 \in \mathbb{N}:\ & (\textit{InputRange}(l_0, l_1, y)\land \lnot \textit{Smaller}(n, l_0) \\
&\land \lnot \textit{Greater}(n, l_1))\Leftrightarrow \textit{Processable}(n, y)
\end{aligned}
\end{equation}
Thereby, $y$ = \textit{Shor-15-Qiskit} in the set of implementations $\mathcal{I}$ and $n$ is the input value, e.g. $n=9$.
\textit{InputRange}$(l_0, l_1, y)$ describes the range of processable input data, for example, $l_0=0$ and $l_1=15$ describe the lower and upper bound.
$\lnot \textit{Smaller}(n, l_0)$ defines that "$n \geq l_0$" and $\lnot \textit{Greater}(n, l_1)$ defines that "$n \leq l_1$", such that $n$ has to be between 0 and 15.
This is all true, if and only if \textit{Processable}$(n, y)$ is true and, thus, \textit{Shor-15-Qiskit} can process $n$.
The selection rule is implementation-specific, and therefore, has to be defined by the developer.
All implementations that can process $n$ are considered in the next phase.

In the \textit{(3) Quantum Computer Analysis \& Selection} phase, the width and depth of the implementations are analyzed and compared with the number of provided qubits and the estimated maximum number of sequential gate layers of the available quantum computers.
The width and depth of the implementation with the given input are determined by using the hardware-specific transpiler.
The determination of width and depth can be realized by plug-ins to support the extensibility for further criteria.
Additionally, the SDKs used by the implementations and supported by the quantum computers are considered.
The general rule for selecting a suitable quantum computer for a particular implementation is defined as follows:
\begin{equation}
\begin{aligned}
\forall x \in \mathcal{Q}, \forall y \in \mathcal{R} \subseteq \mathcal{I}, &\exists s \in \mathcal{S}, \exists q_0, q_1,d_0, d_1 \in \mathbb{N}:\\
&(\textit{Qubits}(q_0, x) \land \textit{Qubits}(q_1, y) \land \lnot \textit{Smaller}(q_0, q_1)\\ &\land \textit{Depth}(d_0, x) \land \textit{Depth}(d_1, y) \land \lnot \textit{Smaller}(d_0, d_1)\\
&\land \textit{Sdk}(s, x) \land \textit{Sdk}(s, y)) \Leftrightarrow \textit{Executable}(y, x)
\end{aligned}
\end{equation}
Thereby, $x$ is a quantum computer of the set of available quantum computers $\mathcal{Q}$, e.g., \textit{IBMQ\_16}, and $y$ is an implementation of the set of remaining implementations $\mathcal{R} \subseteq \mathcal{I}$.
$\textit{Qubits}(q_0, x)$ defines the provided number of qubits $q_0$ of $x$ and $\textit{Qubits}(q_1, y)$ the required number of qubits, or width, $q_1$ of $y$.
$\lnot \textit{Smaller}(q_0, q_1)$ defines that "$q_0 \geq q_1$" to ensure that the quantum computer $x$ does not have less qubits than required by $y$.
$\textit{Depth}(d_0, x)$ defines the maximum depth $d_0$ executable by $x$ and $\textit{Depth}(d_1, y)$ the depth $d_1$ of the transpiled circuit of $y$.
$\lnot \textit{Smaller}(d_0, d_1)$ defines that "$d_0 \geq d_1$", such that the maximum executable depth of the quantum computer $x$ is not smaller than required by the implementation $y$.
Furthermore, the SDK $s \in \mathcal{S}$, e.g. \textit{Qiskit}, used by the implementation, defined by $\textit{Sdk}(y, s)$, must also support the selected quantum computer, defined by $\textit{Sdk}(x, s)$, to ensure their compatibility.
This all is true, if and only if \textit{Executable}$(y, x)$ is true.
In the example in Fig.~\ref{fig:approach} \textit{IBMQ-16} can execute \textit{Shor-15-Qiskit}.

In the \textit{(4) Execution} phase, the selected implementation is finally executed by the selected quantum computer, as seen in Fig.~\ref{fig:approach}.
The \textit{Executor} supports the different SDKs and can be extended by further plug-ins.
Thereby, the required SDK, e.g. \textit{Qiskit}, is used to deliver the quantum circuit to the specific vendor via the cloud.
Eventually, the result is returned and displayed to the user.
\section{Related Work}
For the comparison of different quantum computers, diverse metrics were developed, such as \textit{quantum volume}~\cite{bishop2017quantum} or \textit{total quantum factor~(TQF)}~\cite{Sete2016}.
Additionally, several benchmarks for the quantification of the capabilities of quantum computers were proposed~\cite{Arute2019,benedetti2019}.
However, these metrics only give a rough comparison of the capabilities of the regarded quantum computers and do not consider the aspects of specific quantum algorithms.
Hence, the selection of the quantum computer with the highest score independent of the quantum algorithm and the input data does not always lead to a suitable hardware selection.

Suchara~et~al.~\cite{suchara2013qure} introduce the \textit{QuRE Toolbox}, a framework to estimate the required resources, such as qubits or gates, to execute a quantum algorithm on quantum computers with different physical technologies.
Thereby, the quantum algorithm description is used as input and the required resources for different quantum computers are estimated by QuRE.
Additionally, they consider error-correction and approximate the number of additional gates and qubits that are required to compare the efficiency of different error-correction codes in diverse setups.
However, their focus is on building a suitable quantum computer, not on selecting an existing one.
Therefore, they do not consider the current set of different quantum computers, their supported SDKs, and their limitations.



Also in other domains approaches for decision support exist.
In cloud computing, different approaches for automating the service and provider selection are presented~\cite{han2009efficient,strauch2013decision,zhang2012declarative}.
For \textit{Service Oriented Architecture (SOA)}, decision models are introduced to support the design of application architectures~\cite{olaf1,ZIMMERMANN20091249}.
However, none of these systems include quantum technologies and their special characteristics, such as the limited resources of quantum computers or the varying requirements of quantum algorithms dependent on the input data.


\section{Conclusion \& Future Work}
In this paper, we presented a roadmap for an approach towards a NISQ Analyzer. 
The NISQ Analyzer analyzes and selects for a chosen quantum algorithm and specific input data (i) an appropriate algorithm implementation and (ii) a suitable quantum computer, the implementation can be executed on, by means of defined selection rules.
Thereby, the width and depth of the algorithm implementations are dynamically determined using hardware-specific transpilers and compared with the properties of available quantum computers.
Since quantum computers and implementations are tightly coupled to their SDKs, the compatibility between the SDK used for the implementation and the SDK supported by the quantum computer has to be considered.
The selected implementation is then sent to the corresponding vendor of the selected quantum computer.

In order to automate the selection of an appropriate quantum computer, we are currently implementing the NISQ Analyzer as presented in this paper.
Thereby, the implemented analyzer will be part of a platform for sharing and executing quantum software as proposed in~\cite{leymann2019platform,leymann2020quantum} and currently realized by the project PlanQK\footnote{https://planqk.de}.
We want to analyze the code of the implementations to be able to consider further properties, such as error rates, fidelity, and qubit connectivity. 
Furthermore, we want to remove the dependencies on the hardware-specific transpilers and analyze the width and depth by ourselves. 
In addition, we plan to determine and develop further metrics to enable a more precise analysis and selection of implementations and quantum computers.
%
\section*{Acknowledgements}
\noindent
This work was partially funded by the BMWi project \textit{PlanQK}~(01MK20005N) and the DFG’s Excellence Initiative project \textit{SimTech}~(EXC 2075 - 390740016).

%
%
\bibliographystyle{splncs04}
\bibliography{ms.bbl}
\end{document}